\documentclass[pra,twocolumn,showpacs,amsmath,amssymb]
{revtex4}

\usepackage[dvips]{color}
\usepackage{graphicx}

\begin{document}

\title{State diagrams for harmonically trapped bosons in
optical lattices}
  
\author{Marcos Rigol} 
\affiliation{Department of Physics, Georgetown
University, Washington, DC 20057, USA} 
\author{George G. Batrouni} 
\affiliation{INLN, Universit\'e de Nice-Sophia Antipolis, CNRS;
  1361 route des Lucioles, 06560 Valbonne, France}
\author{Valery G. Rousseau} 
\affiliation{Instituut-Lorentz, LION, Universiteit Leiden, Postbus 9504, 
2300 RA Leiden, The Netherlands} 
\author{Richard T. Scalettar} 
\affiliation{Department of Physics, University of California, Davis, 
California 95616, USA}

\begin{abstract}
We use quantum Monte Carlo simulations to obtain zero-temperature state
diagrams for strongly correlated lattice bosons in one and two
dimensions under the influence of a harmonic confining potential. Since
harmonic traps generate a coexistence of superfluid and Mott insulating
domains, we use local quantities such as the quantum fluctuations of the
density and a local compressibility to identify the phases present in
the inhomogeneous density profiles. We emphasize the use of the
``characteristic density'' to produce a state diagram that is relevant to
experimental optical lattice systems, regardless of the number of bosons
or trap curvature and of the validity of the local-density
approximation. We show that the critical value of $U/t$ at which Mott
insulating domains appear in the trap depends on the filling in the
system, and it is in general greater than the value in the homogeneous
system. Recent experimental results by Spielman {\it et al.} [Phys. Rev.
Lett. {\bf 100}, 120402 (2008)] are analyzed in the context of our
two-dimensional state diagram, and shown to exhibit a value for the
critical point in good agreement with simulations. We also study the
effects of finite, but low ($T\leq t/2$), temperatures. We find that in
two dimensions they have little influence on our zero-temperature
results, while their effect is more pronounced in one dimension.
\end{abstract}


\maketitle

\section{Introduction \label{seci}}

A great amount of experimental and theoretical work \cite{bloch08} has
followed the successful realization of the superfluid(SF)-to-Mott-insulator
transition in ultracold bosonic gases trapped in optical lattices, in
three \cite{greiner02}, two \cite{spielman0708}, and one
\cite{stoferle04} dimensions.  Ultracold atoms on optical lattices are
envisioned as ideal analog simulators of Hamiltonians such as the fermion
Hubbard model, which so far suffers from the lack of reliable analytical
and numerical results.  As a first step toward this goal, intensive
efforts are currently under way to validate this approach by comparing
experimental and theoretical results for systems such as the Bose-Hubbard
model that are amenable to both treatments.  However, even this is
challenging, since the phase diagram of the Bose-Hubbard model is
complicated by issues such as spatial inhomogeneities (recognized early in
Refs.\ \cite{batrouni02,kashurnikov02}), finite-temperature effects
\cite{demarco05,pupillo06,ho07,gerbier07}, and the limited set of 
experimental tools available to characterize those systems.

The phase diagram of the homogeneous Bose-Hubbard model is known to
consist of (i) superfluid phases for all incommensurate fillings and
arbitrary values of the ratio between the on-site repulsion and the
hopping parameter ($U/t$).  The system is also superfluid for
commensurate fillings when $U/t$ is smaller than some critical value
$(U/t)_c$, which depends on the dimensionality of the system and on the
(integer) filling. (ii) Mott insulating phases are present 
for commensurate fillings when $U/t>(U/t)_c$
\cite{fisher89,batrouni90,freericks96,kuhner98,sansone08}. These two
phases have been shown to coexist when a confining potential is added to
the model \cite{jaksch98,batrouni02,kashurnikov02,kollath04,wessel04}. A
feature in the trap, which is advantageous from the experimental
point of view, is that in inhomogeneous systems Mott insulating domains
appear for a broad range of fillings, as opposed to the few commensurate
fillings required for the translationally invariant system.  This
feature comes at a price, sometimes ignored for the sake of simplicity
by both experimentalists and theoreticians working on these systems. In
trapped lattice bosons the critical value $(U/t)_c^T$ for the formation
of Mott insulating domains in different places in the trap not only
depends on the local filling and the dimensionality of the system (the
case for homogeneous systems) but also on the total filling $N$ in the
trap and the curvature of the confining potential $V$.
A useful, but approximate, understanding of the effect of confinement,
which we will not employ here, can be obtained through the use of the
local-density approximation (LDA).

One may think that having to deal with different fillings and trapping
potential parameters in different experiments makes the determination of
a state diagram \cite{foot1} in one particular experimental setup not
relevant to any other.  This is not the case. In Refs.\
\cite{rigol03,rigol04}, it was shown by means of quantum Monte Carlo
(QMC) simulations that for lattice fermions in one dimension one can
define a scaled dimensionless variable, the characteristic density
$\tilde{\rho}=Na(V/t)^{1/2}$ (where $a$ is the lattice spacing), which
allows one to build a state diagram in the plane $\tilde{\rho}$ vs $U/t$
that is insensitive to the individual values of $N$ and $V$ \cite{Others1D}. 
The form for $\tilde \rho$ can most simply be understood as arising from
dimensional arguments: the trap curvature $V$ has units of
energy/length$^2$, so that $(V/t)^{1/2}$ has units of inverse length.
One can then define a length scale $\xi=(V/t)^{-1/2}$, which for trapped
systems plays a role similar to the system size $L$ in the homogeneous
case. The characteristic density $\tilde \rho=N_b a/\xi$ is then a
dimensionless quantity, which for trapped systems is the analog of the
filling per site $n=N_b a/L$ in the homogeneous case.  In other words,
$\tilde \rho$ defines how one should approach the thermodynamic limit in
trapped systems.

The characteristic density is not only relevant to one-dimensional
systems.  It can be generalized to higher dimensions $d$,
$\tilde{\rho}=Na^d(V/t)^{d/2}$ \cite{rigol04a,batrouni08}. In recent work, 
the state diagram of the three-dimensional Fermi-Hubbard model in a harmonic
trap was obtained by using a combination of dynamical mean-field theory
(a treatment which ignores the momentum dependence of the self-energy
while retaining its time fluctuations) and the LDA \cite{leo08}.

One may construct the state diagram in a trap using results for the 
homogeneous system combined with the LDA.  However, this approach is 
only approximate for finite systems. While LDA has been shown to 
give reasonably accurate results in many regimes under a confining potential, 
it is certainly bound to fail close to the critical values of $U/t$ at 
which Mott insulating domains are formed. This is because in the homogeneous 
system, there are diverging correlations when one crosses the quantum critical 
region, i.e., finite-size effects in the trap not only become relevant 
but are also unavoidable. We will show in this paper that in those regimes 
where LDA fails, our exact QMC-based state diagram is an accurate tool 
for characterization of the experimental results.

Since state diagrams are not available for lattice bosons, in this paper
we use world-line quantum Monte Carlo simulations to generate the 
zero-temperature state diagram for the one- and two-dimensional (2D) Bose-Hubbard
model under the influence of a harmonic confining potential. The state
diagram in two dimensions allows us to analyze recent experimental
results for confined two-dimensional bosons in optical lattices
\cite{spielman0708} without the need to perform simulations for the
same parameters as in the experiments. We find that the critical values
for the formation of Mott insulating domains reported in Ref.\
\cite{spielman0708} are consistent (within experimental errors) with our
theoretical results for inhomogeneous systems. Finally, we will discuss
the effect that a small increase in the temperature ($T\leq t/2$) has on
our ground-state results. The latter is found to have little consequence
in two dimensions, where the $U/t$ values which demark the boundaries
between states only change by a few percent.\\

In one dimension, we find that the trapping potential has an even
stronger influence on the critical values of $U/t$ at which Mott
insulating domains appear in the trap. We will also show that low
temperatures have a pronounced effect on the inhomogeneous states in the
trap. The important issue of how to detect experimentally the different
phases in the trap is addressed in several recent works (see, {\it e.g.},
\cite{kollath04,wessel04,gerbier05,sengupta05,rigol06,kollath06,dao07,roscilde08,leo08,delande09})
and will not be discussed here.

\section{Trapped bosons in two dimensions \label{secii}}

\subsection{Hamiltonian and local observables}

For deep enough optical lattices and low temperatures, confined
bosons in two dimensions can be described by the Bose-Hubbard
Hamiltonian \cite{jaksch98}
\begin{eqnarray}
H &=& -t \sum_{\langle i,j \rangle} \left( a^\dagger_{i} a^{}_{j} + \textrm{H.c.} \right) 
+ \dfrac{U}{2} \sum_i n_{i} \left( n_{i} -1 \right) 
\nonumber \\ &&+ V \sum_{i} r_i^2\ n_{i} \,\,.
\label{HubbB}
\end{eqnarray}
Here $a^\dagger_{i}$ and $a_{i}$ are the creation and annihilation
operators of a boson at site $i$, located at a distance
$r_i=\sqrt{x_i^2+y_i^2}$ from the center of the trap.  $x_i$ and $y_i$
are given in units of the lattice spacing, set to unity in this work.
$n_{i} = a^\dagger_{i} a^{}_{i}$ is the particle number operator. The
on-site interaction parameter is denoted by $U$ ($U>0$), the
nearest-neighbor hopping amplitude is denoted by $t$, and $V$ is the curvature of
the harmonic confining potential.  In experiments on optical lattices,
$t$, $U$, and $V$ can be controlled by changing the intensity of the
laser beams that produce the lattice. One can also control $U$
separately using Feshbach resonances \cite{bloch08}. We performed QMC
simulations in the canonical ensemble, using the world-line algorithm
\cite{hirsch82}.  For our-zero temperature study, we have taken the
inverse temperature to be $\beta \, t=18$, which is sufficient to
obtain ground-state results for the observables considered here (see
discussion in Fig.~\ref{Profiles_DiffTemp}), and for discretizing
imaginary time we have chosen $t \, \Delta \tau=0.1$.

In order to create the state diagram, we will monitor three local
observables. The first one is the density. Plateaus with constant
integer density are the so-called Mott insulating domains
\cite{batrouni02}.  However, since the local density always crosses
integer fillings as the total number of bosons in the trap is
increased, and since close to integer local fillings and for large
enough values of $U/t$ one always sees some kind of shoulder appearing
in the density profiles (see, {\it e.g.}, Figs.~\ref{3D_Density} and
\ref{Profiles_DiffTemp}), identifying the formation of Mott domains
only by means of the density is not accurate \cite{rigol03,rigol04}.

The second local quantity we monitor is the quantum fluctuations of
the density, also referred to as the variance of the density,
\begin{equation}
 \Delta_i=\langle n_i^2\rangle -\langle n_i\rangle^2.
\end{equation} 
As shown in Refs.\ \cite{rigol03,rigol04}, this quantity exhibits a
local minimum for densities closest to integer fillings. In addition,
once a Mott insulating domain forms, the value at the minimum equals
the value of the variance in the Mott insulating phase of an
equivalent homogeneous system, i.e., a translationally invariant system
with exactly the same density and $U/t$ \cite{rigol03,rigol04}. We
will show here that in our finite lattice-boson systems (such as the ones
created experimentally), shoulders with $n\simeq 1 $ can appear 
without the variance of the density on those shoulders attaining the
value in the homogeneous Mott insulator.  That is, those shoulders are
not local Mott insulating domains.

The final quantity that we measure here is the local compressibility
defined in Refs.\ \cite{batrouni02,wessel04},
\begin{equation}
 \kappa_i=\dfrac{\partial n_i}{\partial \mu_i} = \dfrac{1}{\beta}
\left[\left\langle  \left( \int_0^\beta d\tau n_i(\tau)\right)^2\right\rangle -
\left\langle \int_0^\beta d\tau n_i(\tau)\right\rangle^2\right].
\end{equation}
This local compressibility quantifies the response in the on-site
density to a local change of the chemical potential. $\kappa_i$ also
exhibits a minimum around integer fillings, and it behaves,
qualitatively, similarly to $\Delta_i$.  An analogous distinction
between equal-time fluctuations and susceptibilities which include
unequal-time correlations plays an important role in the fermion
Hubbard model, where the local susceptibility carries additional
information about the (Kondo) screening beyond that contained in the
equal-time local moment.

\subsection{Density, variance, and local compressibility \label{iiB}}

\begin{figure}[!t]
\begin{center}
  \includegraphics[width=0.39\textwidth,angle=0]{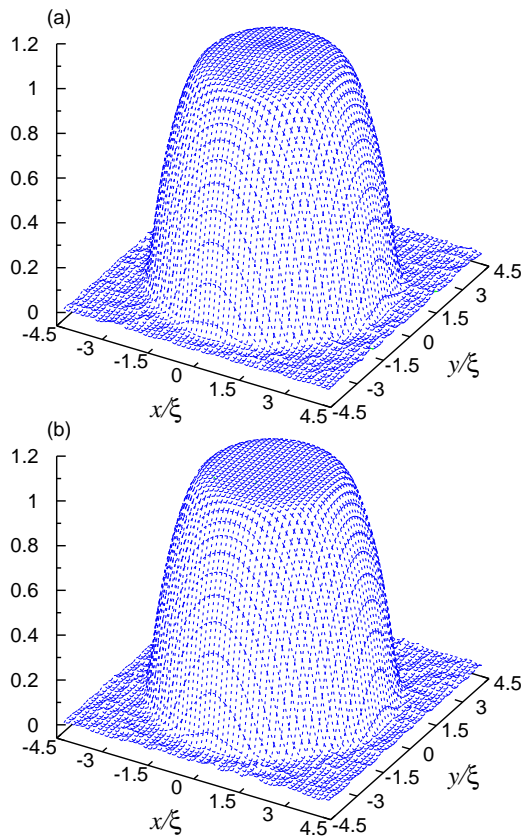}
\end{center}
\vspace{-0.6cm}
\caption{\label{3D_Density} (Color online) Two-dimensional density
profiles for systems with $\tilde{\rho}=30$ ($N_b=1200$ and
$V/t=0.025$), $\beta=18$, $L_x=L_y=60$ (number of lattice sites along
$x$ and $y$), and in-site repulsion (a) $U/t=17.5$ and (b) $U/t=18.5$.
Positions in the trap are normalized by the length scale
$\xi=\sqrt{t/V}$ (see text).  For the homogeneous model (V/t=0) the
critical value $(U/t)_c$ for the SF-Mott transition in $d=2$ is 16.74
\cite{sansone08}.}
\end{figure}

In Fig.~\ref{3D_Density}, we show two density profiles in a 
two-dimensional harmonic trap for values of the on-site repulsive
interaction right before ($U/t=17.5$) and right after ($U/t=18.5$) the
formation of the Mott insulator in the middle of the trap. For
$U/t=17.5$ [Fig.~\ref{3D_Density}(a)], one can see that a shoulder
with density $n\sim 1$ has developed in the system, but the density in
the center of the trap is slightly greater than 1.  For
$U/t=18.5$ [Fig.~\ref{3D_Density}(b)] only a plateau with
density $n=1$ is seen at the central region of the system.  Positions
in the trap are normalized by a length scale $\xi=\sqrt{t/V}$, which
is introduced in Hamiltonian (\ref{HubbB}) by the harmonic
confining potential. With this normalization, systems with different
trapping potentials and fillings, but with the same characteristic
density, have identical density profiles
\cite{rigol03,rigol04,batrouni08}.

\begin{figure}[!t]
\begin{center}
  \includegraphics[width=0.48\textwidth,angle=0]{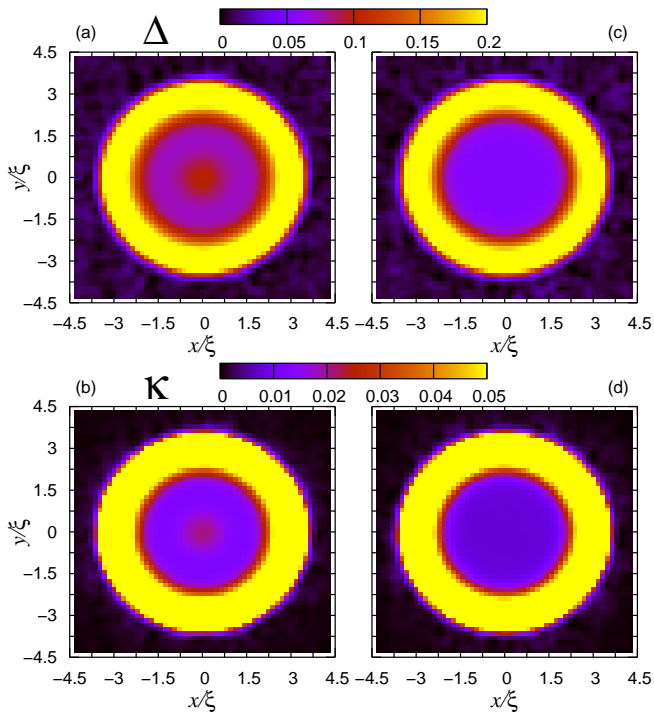}
\end{center}
\vspace{-0.6cm}
\caption{\label{3D_VarianceCompress} (Color online) Intensity plots
of the local quantum fluctuations of the [(a) and (c)] density $\Delta$
and [(b) and(d)] local compressibility $\kappa$ for [(a) and (b)] $U/t=17.5$
and [(c) and (d)] $U/t=18.5$. In these systems $\tilde{\rho}=30$
($N_b=1200$ and $V/t=0.025$), $\beta=18$, and $L_x=L_y=60$.  The
corresponding density profiles are plotted in Fig.~\ref{3D_Density}.
}
\end{figure}

Intensity plots of the variance and local compressibility profiles
corresponding to the systems in Fig.~\ref{3D_Density} are presented
in Fig.~\ref{3D_VarianceCompress}.  This figure shows that both
$\Delta$ and $\kappa$ exhibit minima whenever the density is closest
or equal to 1. In addition, in 
Figs.~\ref{3D_VarianceCompress}(a) and \ref{3D_VarianceCompress}(b)
one can see that the region with $n>1$ at the center of the trap in
Fig.~\ref{3D_Density}(a) is signaled by local maxima of $\Delta$ and
$\kappa$.

\begin{figure}[!t]
\begin{center}
  \includegraphics[width=0.48\textwidth,angle=0]{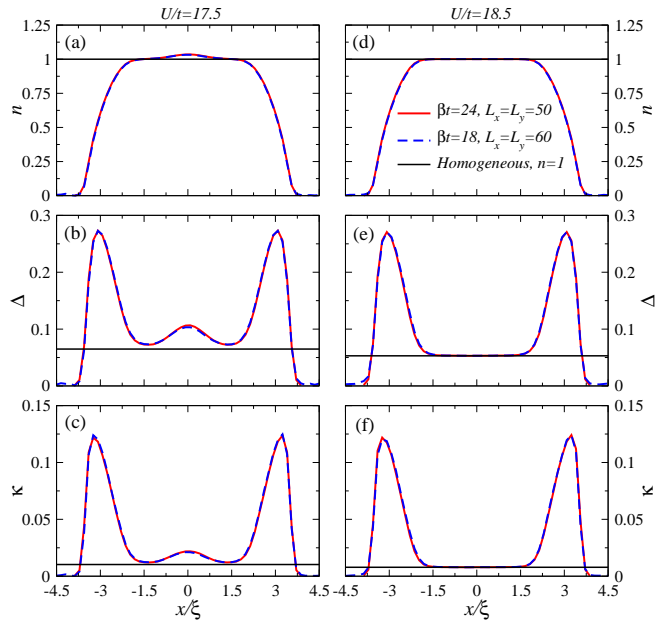}
\end{center}
\vspace{-0.6cm}
\caption{\label{Profiles_DiffTemp} (Color online) Cuts across the
center of the trap for the two-dimensional density, variance, and
local compressibility plots in Figs.~\ref{3D_Density} and
\ref{3D_VarianceCompress} [$\tilde{\rho}=30$ ($N_b=1200$ and
$V/t=0.025$), $\beta=18$, and $L_x=L_y=60$].  Horizontal lines show
the results in homogeneous systems for the same values of $U/t$,
$\beta$, and $n=1$.  These results emphasize that the LDA does not 
hold in some regions of the trap. In the $n\sim
1$ domains in panel (a) that develop around commensurate filling, the
variance in panel (b) does not take on the value predicted by the
homogeneous system for $n=1$. We also plot results for a smaller
system with $L_x=L_y=50$, at lower temperature $\beta=24$, and the
same value of $\tilde{\rho}=30$ ($N_b=1200$ and $V/t=0.025$).}
\end{figure}

A more quantitative understanding of the behaviors of $n,\ \Delta$, and
$\kappa$ can be gained by plotting the changes in these quantities
along one spatial dimension, while the coordinate in the other spatial
dimension is fixed to the center of the trap. The results for the
systems in Figs.~\ref{3D_Density} and \ref{3D_VarianceCompress} are
shown in Fig.~\ref{Profiles_DiffTemp}, which confirms that, indeed,
minima of $\Delta$ and $\kappa$ correspond to regions with $n\simeq 1$
in the density profiles. Along with the results in the trap, we have
plotted as horizontal lines the values of the density, variance, and
compressibility in the Mott insulating phase of the homogeneous system
for exactly the same values of $U/t$.

Figures \ref{Profiles_DiffTemp}(a)--(c) ($U/t=17.5$) show an important
property of these finite trapped systems. While the density profile
can exhibit a shoulder with $n\simeq 1$ [Fig.~\ref{Profiles_DiffTemp}(a)] 
when crossing $n=1$, the variance and local compressibilities on this 
shoulder do not reach their values in the first lobe of homogeneous systems, 
for exactly the same values of $U/t$. The LDA is 
not valid in this region of the trap. Notice that in 
Figs.~\ref{Profiles_DiffTemp}(a)--\ref{Profiles_DiffTemp}(c) such a
shoulder is present for a value of $U/t$ that is greater than the
critical value $(U/t)_c$ for the formation of the Mott insulator in
the homogeneous system, which is $(U/t)_c=16.74$ for $n=1$ in two
dimensions \cite{sansone08}.  Therefore, {\em no} local Mott insulating
phase is found in a trap {\em for} $U/t>(U/t)_c$ and the density at the
center of the trap is $n\geq1$. In an experiment with bosons
trapped in optical lattices, the critical value for the formation of a
Mott insulating state may be greater than the critical value in
homogeneous systems. This is due to the finite curvature of the trapping 
potential and the finite (and sometimes small) extent of the system.
How much the critical value is shifted from the homogeneous prediction 
is something that, as we will show later, will strongly depend 
on the dimensionality of the system. 

In Figs.~\ref{Profiles_DiffTemp}(d)--\ref{Profiles_DiffTemp}(f) one can see 
that for $U/t=18.5$ a full region with $n=1$ occupies the center of the trap. 
In that region, $\Delta$ and $\kappa$ have exactly the same values as that in 
the homogeneous system with $n=1$ for an identical on-site repulsion $U/t$. 
Hence, we conclude that the domain in the middle of the trap is in this case 
a Mott insulator.

To conclude this subsection on local observables, we discuss whether
the lattice size considered in Figs.~\ref{3D_Density} and
\ref{3D_VarianceCompress} is big enough so that {\it boundary conditions}
are irrelevant, and whether the temperature is low enough so that $n,\
\Delta$, and $\kappa$ behave as they will in the ground state. In
Fig.~\ref{Profiles_DiffTemp}, we present results for a system with
identical Hamiltonian parameters and filling but at a lower
temperature $\beta=24$ and in a smaller lattice with $L_x=L_y=50$
sites in each direction.  They can be seen to be essentially
indistinguishable from those with $\beta=18$ and $L_x=L_y=60$.  Hence,
at least for our local observables, the temperature considered is low
enough and the systems sizes are big enough so that they have no
influence in our results.
 
\subsection{State diagram}

Uniform systems of different sizes have identical properties as long as
they are chosen to have the same density $\rho=N_b/L^d$, or in a
lattice the same $n=N_b a^d/L^d$, and are sufficiently large.  In a
similar way, the characteristic density generalizes this for confined
systems and allows one to obtain state diagrams that are independent 
of particular choices of boson number and trap curvature provided the 
systems are also large. Finite systems, such as the ones realized 
experimentally, pose an additional complication as finite-size effects 
become relevant. They are most important close to where local phases 
(such as those described in Sec.\ \ref{iiB}) appear for the 
first time in the trap. This implies that, as also shown in Sec.\ \ref{iiB}, 
the LDA may fail to describe 
the actual density profiles in an ultracold gas experiment on an optical 
lattice.

\begin{figure}[!b]
\begin{center}
  \includegraphics[width=0.48\textwidth,angle=0]{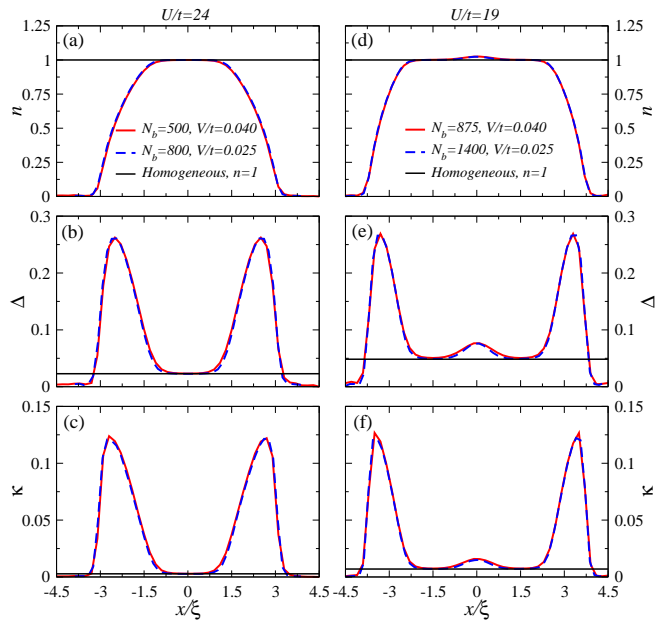}
\end{center}
\vspace{-0.6cm}
\caption{\label{Profiles_DiffFilling} (Color online) Density,
variance, and local compressibility in one direction across the center
of the trap. We compare systems with the same characteristic density
but different trapping potentials and fillings. Left panels [(a)--(c)]:
$U/t=24$, $\tilde{\rho}=20$, with $N_b=800$, $V/t=0.025$ and
$N_b=500$, $V/t=0.04$. Right panels [(d)--(f)]: $U/t=19$,
$\tilde{\rho}=35$, with $N_b=1400$, $V/t=0.025$ and $N_b=875$,
$V/t=0.04$.  In all cases $\beta=18$. Horizontal lines show
the results for homogeneous systems with $n=1$ and the same values of
$U/t$ and $\beta$ as in the trap.  }
\end{figure}

In this section we show that, for systems like the ones realized 
experimentally, a state diagram in the plane $\tilde{\rho}$
vs $U/t$ can accurately predict the local phases in trapped experiments even 
when the LDA fails. What this means is that for those 
system sizes (where the occupied  part of the lattice has $50 \lesssim 
L_x,L_y \lesssim 100$), changing the specific trap parameters does not 
appreciably change the observed local phases under the confining potential. 
In order to create such a state diagram, we have considered systems 
with two different values of the trap curvature and many different fillings.
We fixed the inverse temperature to $\beta=18$ and the largest lattices 
considered here had $L_x=L_y=64$ lattice sites. The lattice sizes 
considered here are comparable to the ones realized experimentally
\cite{spielman0708}.

In the left panels of Fig.~\ref{Profiles_DiffFilling}[(a)--(c)], we
compare two systems with the same characteristic density
$\tilde{\rho}=20$, but two different trapping potentials and fillings
[$N_b=800$, $V/t=0.025$ and $N_b=500$, $V/t=0.04$]. Our results for
the scaled density, variance, and local compressibility profiles are
essentially indistinguishable between those two traps, despite the large
difference in particle number and curvature.  In 
Figs.~\ref{Profiles_DiffFilling}(a)--\ref{Profiles_DiffFilling}(c), we 
have chosen one of the lowest
characteristic densities that support a Mott insulating state in the
trap. With increasing $\tilde \rho$, at the same values of $V/t$, the
differences between the scaled profiles in the two traps become even
smaller.

In the right panels of Fig.~\ref{Profiles_DiffFilling}[(d)--(f)], we
compare two systems in which an additional superfluid phase with $n>1$
is present inside a Mott insulating domain with $n=1$. This state has
a richer structure than the one in the left panel. The characteristic
density is the same in both systems, $\tilde{\rho}=35$, but the trap
curvature and fillings are different ($N_b=1400$, $V/t=0.025$ and
$N_b=875$, $V/t=0.04$). We have chosen the ratio of $U/t$ in this case
to be very close to the critical value for the formation of the Mott
insulator at the center of the trap. Figures
\ref{Profiles_DiffFilling}(d)--\ref{Profiles_DiffFilling}(f) show that 
even in this more complicated case, with several different domains and 
close to a transition between states, the scaled results for our local
observables are also almost indiscernible even for two very
different systems, as long as they have the same characteristic density.

\begin{figure}[!b]
\begin{center}
  \includegraphics[width=0.48\textwidth,angle=0]{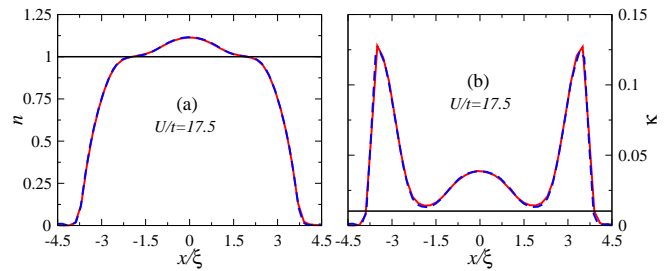}
\end{center}
\vspace{-0.6cm}
\caption{\label{Profiles_WRONGLDA} (Color online) 
(a) Density and (b) local compressibility in one direction across the 
center of the trap. We compare systems with the same characteristic density
$\tilde{\rho}=35$ but different trapping potentials and fillings of
$V/t=0.025$, $N_b=1400$ and $V/t=0.04$, $N_b=875$. Horizontal lines show
the results for homogeneous systems with $n=1$ and the same values of
$U/t$ and $\beta=18$ as in the trap.}
\end{figure}

In Figs.~\ref{Profiles_DiffTemp}(a)--\ref{Profiles_DiffTemp}(c), we have shown 
results for local quantities in a system that cannot be described 
within the LDA. In Fig.~\ref{Profiles_WRONGLDA}, we 
provide an additional example where two different confined systems 
with $U/t>(U/t)_c$ [$(U/t)_c$ from the homogeneous case] fail to 
exhibit Mott insulating domains. This figure also shows that even 
though the LDA is clearly not applicable to those systems, the density 
profiles and local compressibilities in both of them, which have the 
same characteristic density, are almost indistinguishable. As in 
previous figures, the number of particles in the largest 
system is almost twice that of the smallest. These results confirm that
the characteristic density, being the proper quantity to define
the thermodynamic limit in a trapped system, provides genuine guidance
for finite systems when the LDA is not valid.

Following the discussion in Sec.\ \ref{iiB}), we locate Mott 
insulating domains in the trap by comparing the variance of the density 
and local compressibilities in the confined system with those of a Mott
insulator in the homogeneous case. Whenever they coincide, we conclude
that a local Mott insulating domain is present.  With this criterion,
in most cases Mott insulating domains emerge in the trap for the same 
value of $U/t$ independent of whether one considers $\Delta$
or $\kappa$. In a few cases, especially at low fillings, we found
$\Delta$ in the trap to agree with the value in the homogeneous case
before $\kappa$ did, usually by a difference $\delta(U/t)\sim0.5$. In
those cases, the value of $U/t$ reported at the state boundary is the 
average between the one obtained by $\Delta$ and the one obtained by $\kappa$.  
Note that this discrepancy represents, at worst, an uncertainty in the 
boundary of only a few percent.

\begin{figure}[!t]
\begin{center}
  \includegraphics[width=0.43\textwidth,angle=0]{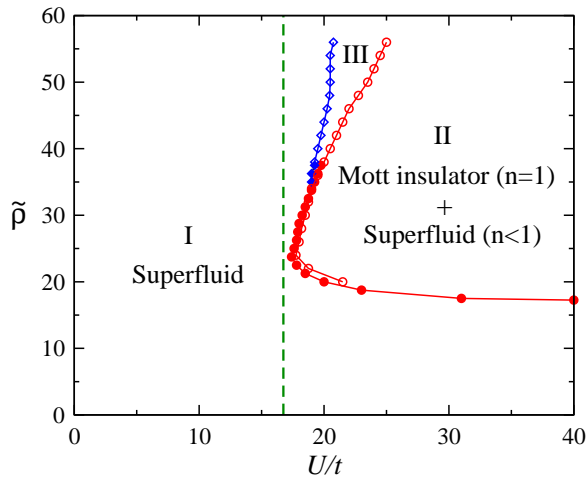}
\end{center}
\vspace{-0.6cm}
\caption{\label{StateDiagram} (Color online) State diagram for bosons
in a two-dimensional lattice in a harmonic confining potential. The
boundaries between states were determined using two different traps,
with $V/t=0.025$ (filled symbols) and $V/t=0.04$ (open symbols).  Both
traps lead to very similar results except for some small differences
at the lowest characteristic densities. The states are: (I) a pure 
superfluid phase [Figs.~\ref{Profiles_DiffTemp}(a)--\ref{Profiles_DiffTemp}(c)], 
(II) a Mott insulating phase at the center of the trap surrounded by a superfluid 
phase with $n<1$ [Figs.~\ref{Profiles_DiffTemp}(d)--\ref{Profiles_DiffTemp}(f)
and Figs.~\ref{Profiles_DiffFilling}(a)--\ref{Profiles_DiffFilling}(c)], 
(III) a superfluid phase with $n>1$ at the center of the trap surrounded 
by a Mott insulating phase with $n=1$, and an outermost superfluid phase with $n<1$ 
[Figs.~\ref{Profiles_DiffFilling}(d)-\ref{Profiles_DiffFilling}(f)]. 
The vertical dashed line signals the critical value of $U/t$ for the 
formation of the Mott insulator with $n=1$ in the homogeneous case 
\cite{sansone08}. The inverse temperature in the trapped systems is 
$\beta=18$. }
\end{figure}

Figure \ref{StateDiagram} is the central result of this paper: the
state diagram for lattice bosons in a two-dimensional lattice under
the influence of a harmonic trap.  The different states depicted
represent the following. (I) Only a superfluid phase is present in the trap, the
situation in Figs.~\ref{Profiles_DiffTemp}(a)--\ref{Profiles_DiffTemp}(c). 
(Notice that in this
state the density at the center of the trap can take arbitrarily high
values.)  (II) A Mott insulating phase at the center of the trap is
surrounded by a superfluid phase with $n<1$. Examples of this state
are given in Figs.~\ref{Profiles_DiffTemp}(d)--\ref{Profiles_DiffTemp}(f) and 
Figs.~\ref{Profiles_DiffFilling}(a)-\ref{Profiles_DiffFilling}(c). 
(III) A local superfluid phase with
$n>1$ is present at the center of the trap. This phase is
surrounded by a Mott insulating domain with $n=1$, inside a superfluid
phase with $n<1$. An example of that state can be found in the right
panel of Fig.~\ref{Profiles_DiffFilling}(d)--\ref{Profiles_DiffFilling}(f).

An important feature of the state diagram of Fig.\ \ref{StateDiagram} 
is that except for a small window of characteristic densities
($\tilde{\rho}\sim 23$), where the critical value of $U/t$ for the
formation of the Mott insulating domain at the center of the trap
$(U/t)_c^t$ is very close to the value in homogeneous systems
$(U/t)_c$, for most characteristic densities $(U/t)_c^t$ departs
significantly from $(U/t)_c$; i.e., this state boundary depends
strongly on $\tilde{\rho}$. At least for fermions \cite{jordens08},
this is the state boundary that currently can be accurately determined
experimentally. This is because the total double occupancy, which 
is a good indicator for distinguishing states with a Mott insulator at 
the center of the trap from states with higher density \cite{leo08}, 
can be accurately measured in such systems \cite{jordens08}. 
Similar measurements in bosonic systems, which have site occupancy 
higher than two, could allow experimentalists to reproduce our 
results over an even broader range of the state diagram.

The other state boundary of interest is the one between states I and
III.  This boundary, on the other hand, does not depend strongly on
the characteristic density \cite{foot2}. This means that by experimentally 
measuring the boundary between states I and III, which is sensitive to 
transport and coherence probes, one would obtain a critical value of 
$U/t$ for the formation of the Mott insulator that is close to the 
one in the homogeneous system.

To conclude, we should add that with increasing characteristic density
new states with Mott insulating domains with $n>1$ appear in the
trap.  Those states will form for larger values of the in-site
repulsive interaction, and will be more difficult to detect
experimentally.

\subsection{Comparison with NIST experimental results}

As mentioned in Sec.\ \ref{seci}, recent experiments have achieved the
superfluid-to-Mott-insulator transition in two dimensions
\cite{spielman0708}.  Here we provide a quantitative comparison with
our state diagram.

We use the precise values of the curvature $V$ generated by the
optical and magnetic traps, and interaction and tunneling energies $U$
and $t$ (computed using a band structure calculation) appropriate to
the NIST experiments \cite{thanks}.  The final parameter of interest
is the filling.  The NIST experiments employed an array of about 70
independent 2D systems of different filling.  The number of particles
reported for the central 2D slice (the one with maximal filling)
\cite{spielman0708} was $N_b\approx 3000$, while the average was
$N_b\approx 2000$.

\begin{figure}[!t]
\begin{center}
  \includegraphics[width=0.43\textwidth,angle=0]{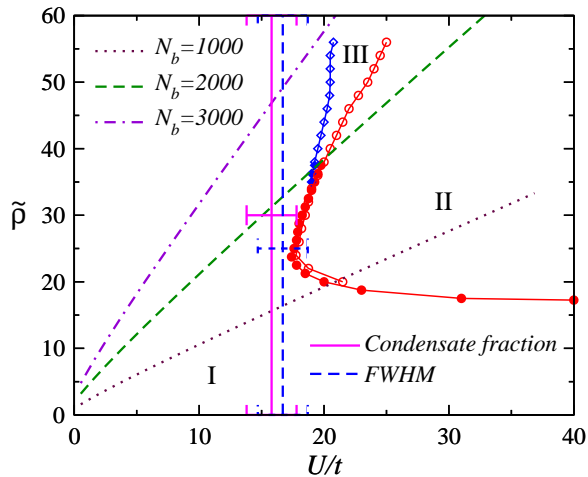}
\end{center}
\vspace{-0.6cm}
\caption{\label{StateDiagramExperiments} (Color online) State diagram
in Fig.~\ref{StateDiagram} with experimental trajectories for systems
with different fillings, $N_b=1000,\ 2000,$ and 3000. $V$, $t$, and
$U$ have been chosen to correspond to the specific experiments is
Ref.\ \cite{spielman0708}. Vertical lines depict the critical values
of $U/t$ reported in Ref.\ \cite{spielman0708}, computed using the
condensate fraction in one case and the full width at half maximum
(FWHM) of the momentum distribution function in the other
\cite{spielman0708}. The experimental errors are also plotted.  The
experiment and QMC give values for the critical coupling in excellent
agreement.  }
\end{figure}

Figure \ref{StateDiagramExperiments} shows the trajectories of systems
with different fillings, and the NIST trapping and energy parameters,
in the QMC state diagram.  We have also plotted the experimentally
reported critical values of $U/t$ for the formation of the Mott
insulator obtained using the condensate fraction and the 
FWHM of the momentum distribution function.  For
intermediate fillings ($N_b$ between 1000 and 2000 particles)
corresponding to the average filling of the array of 2D layers, the
experimental results intercept (within their error bars) the QMC
boundary between states I and II. The error bars reported in the
experiments arise mainly from the uncertainty in the lattice depth,
which translates into around a $\pm 10\%$ uncertainty in the ratio
$U/t$.  For the largest fillings of the 2D slices ($N_b$ between 2000
and 3000 particles), we find that the experimentally reported critical
values are very close to the boundary between states I and III.

Overall, the agreement between our numerical calculations and the
experimental results is remarkable. It could even be considered 
surprising, taking into account that in the experiments only global 
quantities such as the condensate fraction and the FWHM of the momentum 
distribution function have been measured, while our state diagram 
is based on the behavior of local quantities such as the variance of 
the density and the local compressibility. The agreement between 
those two approaches is a priori not guaranteed \cite{wessel04}.
New experiments in which the filling in the 2D sections is better 
controlled could allow experimentalists to map the state diagram 
in Fig.~\ref{StateDiagram}. Given the presence of the trapping 
potential in the experiments and the fact LDA fails close to 
$(U/t)_c$ of the homogeneous system, we believe this is a better 
strategy to follow to validate experiments than trying to map the 
phase diagram of the homogeneous system. Recent progress on the 
direct imaging of the density profile will also allow further 
detailed comparisons with simulation.

\subsection{Finite temperature}

In order to better compare with the experiments, it is important
to understand how a small increase in the temperature affects the
zero-temperature state diagram reported in this paper. 
By small we mean a temperature that is low compared to the value of
the hopping parameter, so we consider here temperatures $T\leq t/2$. 
The effect of higher temperatures ($t<T<U$) on the Mott insulating state 
in trapped systems has been discussed in previous works 
(see, {\it e.g.}, Refs.\ \cite{demarco05,pupillo06,ho07,gerbier07}).

\begin{figure}[!b]
\begin{center}
  \includegraphics[width=0.48\textwidth,angle=0]{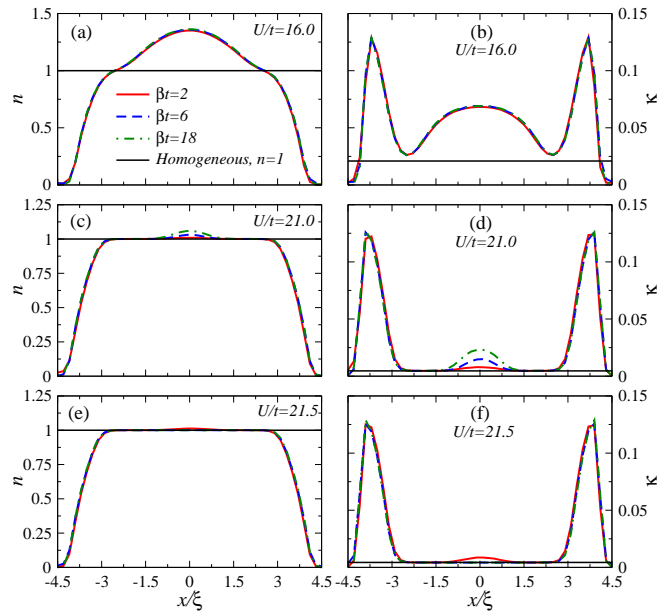}
\end{center}
\vspace{-0.6cm}
\caption{\label{Profiles_FiniteTemperature} (Color online) Comparison
between local quantities in systems with different temperatures. We
have plotted the density (left panels) and local compressibility
(right panels) in one direction across the center of the trap. The
ratio between the on-site repulsion and the hopping parameter is
increasing from top to bottom:[(a) and (b)]$U/t=16$, [(c) and (d)] $U/t=21$,
and [(e) and (f)] $U/t=21.5$. All systems have the same characteristic density
$\tilde{\rho}=44$ ($N_b=1100$, $V/t=0.04$).  Horizontal lines show the
results for homogeneous systems with $n=1$, and the same values of
$U/t$ as in the trap. The differences bewteen $\kappa$ in homogeneous
systems with $\beta t=2,\ 6,$ and 18 are indistinguishable in the
figure.}
\end{figure}

Our QMC simulations for two-dimensional systems with $T\leq t/2$ show 
that low temperatures have a small effect on the state diagram in 
Fig.~\ref{StateDiagram}. We have found that in general the state 
boundaries move to larger values of $U/t$, with a change of up to
4\% of the ground-state value depending on the filling. This effect 
is particularly evident for the lowest characteristic densities. For larger 
characteristic densities, the ones that support state III in 
Fig.~\ref{StateDiagram}, we find the effect of low temperatures to be 
less important.

As an example of our results, in Fig.~\ref{Profiles_FiniteTemperature} 
we compare density profiles and compressibilities in confined systems 
with $\tilde{\rho}=44$ at different temperatures and interaction strengths. 
One can see there that while density and compressibility profiles in 
state I are almost indistinguishable for the temperatures considered here
[Figs.~\ref{Profiles_FiniteTemperature}(a) and \ref{Profiles_FiniteTemperature}(b), 
where $U/t=16$], the ones in state III can exhibit apparent changes when close 
to the boundary with state II [Figs.~\ref{Profiles_FiniteTemperature}(c) 
and \ref{Profiles_FiniteTemperature}(d), where $U/t=21$]. However, with just 
a small increment of the interaction strength so that one creates state II 
in the ground state [Figs.~\ref{Profiles_FiniteTemperature}(e) and 
\ref{Profiles_FiniteTemperature}(f), where $U/t=21.5$] 
the system has a full Mott insulating domain at the center of the trap 
for $\beta=18$ and $\beta=6$, while $\beta=2$ is very close to forming one.
The generic effect of the temperature can be seen to be a displacement 
of the state boundaries to larger values of $U/t$.

\section{Trapped bosons in one dimension}

\begin{figure}[!htb]
\begin{center}
  \includegraphics[width=0.43\textwidth,angle=0]{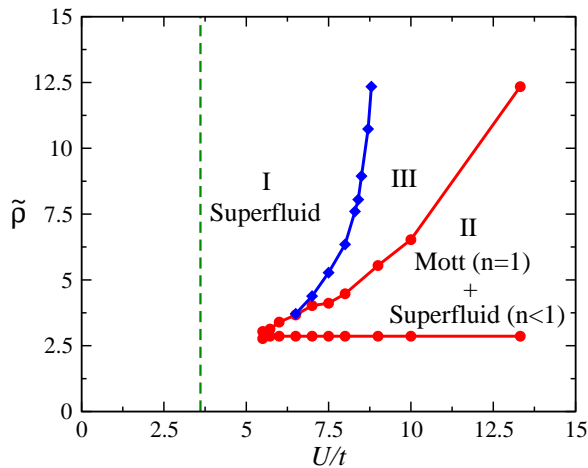}
\end{center}
\vspace{-0.6cm}
\caption{\label{StateDiagram1D} (Color online) State diagram for
bosons in a one-dimensional lattice with a harmonic confining
potential.  The state diagram was built using a trap with
$V/t=0.008$, $L=100$, and $\beta=10$. The states are designated in the 
same way as in the two-dimensional case: (I) a pure superfluid phase, 
(II) a Mott insulating phase at the center of the trap surrounded by a 
superfluid phase with $n<1$, (III) a superfluid phase with $n>1$ at the 
center of the trap surrounded by a Mott insulating phase with $n=1$, 
and an outermost superfluid phase with $n<1$. The vertical dashed line 
signals the critical value of $U/t$ for the formation of the Mott insulator 
with $n=1$ in the homogeneous case \cite{kuhner98}.}
\end{figure}

In one dimension, previous work has already shown that systems with
the same characteristic density $\tilde{\rho}=N_b\sqrt{V/t}$ have
identical re-scaled density profiles \cite{batrouni08}, so here we
will not present a detailed discussion (i.e., the $d=1$ analogs of
Figs.\ \ref{Profiles_DiffTemp}, \ref{Profiles_DiffFilling}, and
\ref{Profiles_WRONGLDA} presented for $d=2$). In addition, the behavior 
of the local observables defined in Sect.\ \ref{secii} is qualitatively 
similar to that in two-dimensional systems. Therefore, in this section, 
we use the same criteria to create the $d=1$ state diagram, 
depicted in Fig.~\ref{StateDiagram1D}. The labeling of the states 
follows the same convention as in two dimensions.

Figure \ref{StateDiagram1D} shows that in one dimension the trapping 
potential has a more pronounced effect on displacing the critical 
value for the formation of the Mott insulator toward larger values of
$U/t$. This can be understood if one compares the homogeneous Mott lobes in one 
dimension with those of higher dimensional systems, and is somehow similar to 
what was found in Refs.\ \cite{rigol03,rigol04} for the fermionic case, where, 
in the trap, the lowest value of $U/t$ at which 
a local Mott insulating phase was found was $U/t\sim 3$, to be compared with 
$(U/t)_c=0$ in the homogeneous case. Another important difference between
the state diagrams in two and one dimensions is that in the latter
all states boundaries have a stronger dependence on the characteristic 
density. 

In order to study the effect of finite temperatures in one-dimensional systems, 
we performed quantum Monte Carlo simulations using the stochastic Green's-function 
algorithm \cite{rousseau08}. This algorithm allows us to compute the momentum
distribution function of the bosons, a quantity that is very difficult to compute 
with the world-line approach. 

\begin{figure}[!t]
\begin{center}
  \includegraphics[width=0.48\textwidth,angle=0]{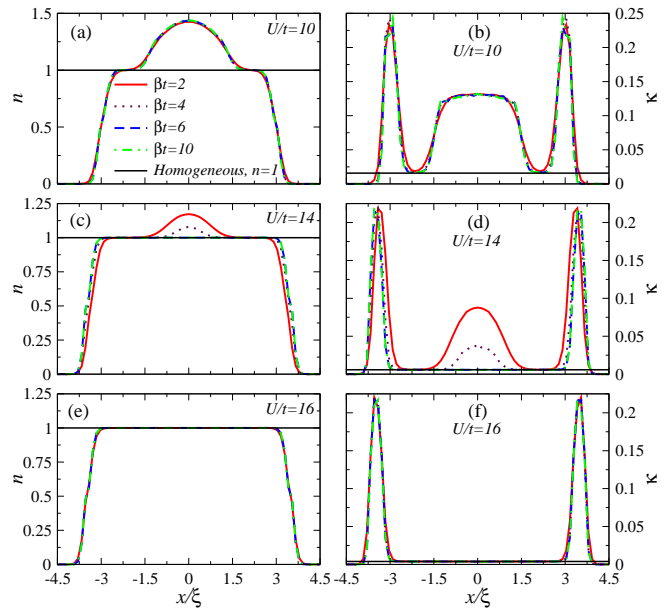}
\end{center}
\vspace{-0.6cm}
\caption{\label{Profiles_FiniteTemperature1D} (Color online) Comparison
between local quantities [density (left panels) and local compressibility
(right panels)] in one-dimensional systems with different temperatures. 
The ratio between the on-site repulsion and the hopping parameter is
increasing from top to bottom: [(a) and (b)] $U/t=10$, [(c) and (d)] $U/t=14$,
and [(e),(f)] $U/t=16$. All systems have the same characteristic density
$\tilde{\rho}=6.96$ ($N_b=55$, $V/t=0.016$).  Horizontal lines show the
results for homogeneous systems with $n=1$, and the same values of
$U/t$ as in the trap. The differences between $\kappa$ in homogeneous
systems with $\beta t=2,\ 4,\ 6,$ and 10 are indistinguishable in the
figure.}
\end{figure}

In Fig.~\ref{Profiles_FiniteTemperature1D}, we show examples of the 
effect of the temperature in one dimension. The temperature 
is increased up to $T=t/2$, similarly as in the two-dimensional case. 
Figures \ref{Profiles_FiniteTemperature1D}(a) and \ref{Profiles_FiniteTemperature1D}(b) 
depict a system that at zero temperature is in state III in Fig.~\ref{StateDiagram1D}.
Increasing the temperature from $\beta=10$ to $\beta=2$ in that state 
does not produce large changes in the density and compressibility profiles.
The region with $n\sim 1$ is the most affected by the temperature, 
as the Mott plateaus to the side of the central region with $n>1$ have shrunken. 

Figures \ref{Profiles_FiniteTemperature1D}(c) and \ref{Profiles_FiniteTemperature1D}(f) 
deal with the state that at zero temperature has a Mott insulator at the center 
of the trap (state II in Fig.~\ref{StateDiagram1D}). For $U/t=14$
[Figs.~\ref{Profiles_FiniteTemperature1D}(c) and \ref{Profiles_FiniteTemperature1D}(d)], 
one can see that increasing the temperature beyond $\beta=6$ melts the Mott insulator at
the center of the trap, producing a central region with $n>1$. As one
increases $U/t$, $U/t=16$
[Figs.~\ref{Profiles_FiniteTemperature1D}(e) and \ref{Profiles_FiniteTemperature1D}(f)], 
one can see that the central Mott insulating domain survives for all temperatures 
analyzed here. 

Overall, Fig.~\ref{Profiles_FiniteTemperature1D} shows that the effect
of low temperatures in the state diagram in one dimension is
qualitatively similar to that in two dimensions; i.e., the boundaries of
state II move to larger values of $U/t$. However, our simulations show
that one--dimensional systems are more affected by finite temperatures
than their two-dimensional counterparts, so that any experimental
comparison with our state diagram in Fig.~\ref{StateDiagram1D} would
require very low temperatures. These results also mean that the critical
values for the formation of the insulator in one-dimensional trapped
systems at finite temperatures are further away from the critical value
in the homogeneous system at zero temperature.

\begin{figure}[!b]
\begin{center}
  \includegraphics[width=0.48\textwidth,angle=0]{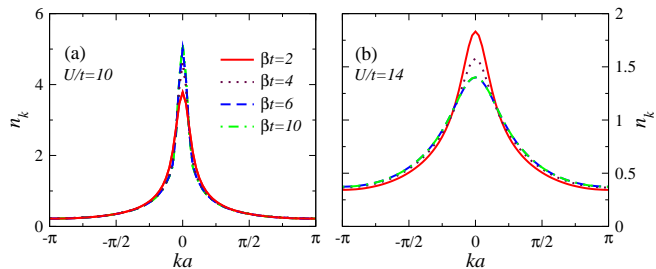}
\end{center}
\vspace{-0.6cm}
\caption{\label{nk_FiniteTemperature1D} (Color online) Momentum distribution
function in one-dimensional systems with different temperatures. 
The ratio between the on-site repulsion and the hopping parameter are
(a) $U/t=10$ (at zero temperature the system is in state III), and 
(b) $U/t=14$ (at zero temperature the system is in state II). 
All systems have the same characteristic density 
$\tilde{\rho}=6.96$ ($N_b=55$, $V/t=0.016$).}
\end{figure}

To conclude, we show in Fig.~\ref{nk_FiniteTemperature1D} the effect of
the temperature on the momentum distribution function of the systems
with $U/t=10$ and $U/t=14$ in Fig.~\ref{Profiles_FiniteTemperature1D}.
(The momentum distribution function for the systems with $U/t=16$
remains almost unchanged for the temperatures considered here.) Figure
\ref{nk_FiniteTemperature1D}(a) shows that when most of the system is in
a superfluid phase at zero temperature, increasing the temperature leads
to a reduction in the zero-momentum peak occupation, which is a
consequence of the reduction in the correlation length in the superfluid
domains. On the other hand, Fig.~\ref{nk_FiniteTemperature1D}(b) shows
that when the ground state of the system has a very large Mott
insulating domain at the center of the trap, an increase in the
temperature can increase the occupation of the zero-momentum peak when
the Mott insulator melts to give rise to a region with $n>1$ at the
center of the trap.

\section{Conclusions}

In this paper we have computed the state diagrams for the formation of
Mott insulating domains with $n=1$ in lattice-boson systems confined
by harmonic traps. Results have been presented in one and two
dimensions and the lattice sizes considered had between 50 and 100 sites
in each direction, values which are comparable to those studied by the 
NIST group.

We have shown that for the system sizes considered, state boundaries 
are accurately determined by the characteristic density, providing a 
state diagram for different experimental choices of trap curvature 
and particle number. Key features of these state diagrams are that the 
critical values of $U/t$ for the formation of a Mott insulator in a 
trap measured experimentally need not be the same as in homogeneous 
systems and that the LDA does not hold close to 
the state boundaries.
We find that in two dimensions, the lowest value of $U/t$ that supports 
a local insulator in the trap is $U/t=17.4$, which is approximately 
4\% greater than 
the critical value in the homogeneous case, $(U/t)_c=16.74$. We have also 
shown that $(U/t)_c^T$ for the formation of the Mott insulator at the 
center of the trap increases with the characteristic density. These results 
could be verified experimentally by measuring double or higher occupancy 
in bosonic systems.  

A comparison of our two-dimensional results with the ones of experiments 
at NIST shows that the particular trap, lattice depth, and particle 
filling used there give trajectories in the state diagram which intersect 
the critical coupling line near its extremal value of $U/t$. The 
transition point reported is in good agreement with our QMC simulations, 
which explicitly include the confining potential. We have also shown that, 
in two dimensions, finite (but low, $T\leq t/2$) temperatures have little 
effect on our ground-state results.

In one dimension, we find that the trapping potential has a stronger 
effect in moving the critical values for the formation of the Mott insulator 
toward larger values of $U/t$. Actually, the lowest value of $U/t$ at 
which we find a local Mott insulating phase in the trap (with up to
100 sites) is $U/t=5.5$, that is, a 52\% increase over the critical value 
in the homogeneous case, $(U/t)_c=3.61$. We also find that state boundaries 
in one dimension are more sensitive to the characteristic density in the 
trap and to low temperatures than those in two-dimensional systems.

\begin{acknowledgments}
M.R. acknowledges support from startup funds from Georgetown
University and from the US Office of Naval Research. G.G.B. was supported by 
the CNRS (France) Grant No.\ PICS 3659. The work of R.T.S. was supported under 
USARO Award No.\ W911NF0710576 with funds from the DARPA OLE Program. V.G.R. 
was supported by the research program of the Stichting voor Fundamenteel 
Onderzoek of der Materie (FOM) with funds from the Nederlandse 
Organisatie voor Wetenschappelijk Onderzoek (NWO). We would like 
to thank J. V. Porto and I. B. Spielman for useful discussions and 
for providing us with their experimental data. We are also grateful 
to B. H. Amabo for his guidance and support.
\end{acknowledgments}


\begin{thebibliography}{99}
\bibitem{foot1}
Where each state can, in general, be composed by superfluid and 
insulating domains with different densities.
\bibitem{Others1D} For homogeneous systems, the Fermi-Hubbard Hamiltonian 
in one dimension can be exactly solved by Bethe ansatz. This, in 
combination with the LDA, has allowed 
different groups to reproduce (up to small differences at the state 
boundaries \cite{xianlong06}) and extend the QMC state diagram under 
a harmonic confinement \cite{liu05,campo05,heiselberg06}.
\bibitem{foot2}
In the thermodynamic limit, based on the LDA, one 
would expect this boundary to occur at constant $U/t$ starting from the tip 
in the boundary of state II.
\end{thebibliography}
\end{document}